\documentclass[twocolumn,preprintnumbers,amsmath,amssymb]{revtex4}
\input{epsf}

\usepackage{dcolumn}
\usepackage{bm}

\def\lsim{\,{}_\sim^{<}\,}

\begin{document}

\title{Weak Primordial Magnetic Fields and Anisotropies in the Cosmic Microwave Background Radiation}

\author{Karsten Jedamzik$^1$ and Tom Abel$^{2,3,4}$}

\affiliation{
$^1$Laboratoire de Univers et Particules, UMR5299-CNRS, Universit\'e
de Montpellier II, F-34095 Montpellier, France \\
$^{2}$Kavli Institute for Particle Astrophysics and Cosmology, SLAC/Stanford University, 2575 Sand Hill Road, Menlo Park, CA 94025, USA\\
${^3}$Zentrum f\"ur Astronomie der Universit\"at Heidelberg, Institut f\"ur Theoretische Astrophysik, Albert-Ueberle-Str. 2, 69120 Heidelberg, Germany\\
${^4}$Heidelberg Institut f\"ur Theoretische Studien, Schloss-Wolfsbrunnenweg 35, 69118 Heidelberg, Germany}

\begin{abstract}
It is shown that small-scale magnetic fields present before recombination
induce baryonic density inhomogeneities of appreciable magnitude. The presence
of such inhomogeneities changes the ionization history of the Universe, which
in turn decreases the angular scale of the Doppler peaks and increases
Silk damping by photon diffusion. This unique signature could be used to  
(dis)prove the existence of primordial magnetic fields of strength
as small as $B\simeq 10^{-11}\,$Gauss by upcoming cosmic microwave 
background observations. 
\end{abstract}


\maketitle


Primordial magnetic fields may well have been generated during early
cosmic phase transitions, during an inflationary epoch (in case conformal
invariance is broken), or during an epoch of baryogenesis,
among other. In fact, it seems unlikely that the early Universe was not
magnetized due to the multitude of possibilities. The question is rather 
of which strength such fields would be, and on what typical length scales 
they would reside. Recently, it has been claimed that the surprisingly
weak flux of GeV $\gamma$-rays in the direction of three TeV-blazars
may be best understood by the presence of cosmic magnetic fields 
of relatively weak strength~\cite{Neronov} filling
a large fraction of 
space~\cite{Dolag}(see Ref.~\cite{plasma} however). 
Such fields could explain why
secondary GeV $\gamma$-rays, induced by TeV $\gamma$-rays pair producing
on the infrared background, with the produced $e^{\pm}$ subsequently
inverse Compton scattering on the cosmic microwave background radiation
(CMBR), would be moved out of the light
cone due to the curved trajectories of the $e^{\pm}$. 

It would be important to find other observational signatures of such
putative primordial magnetic fields. A prime candidate here are precision
observations of anisotropies of the CMBR. A larger number of studies have
been presented, with the majority of studies assuming substantial
fields on 10-100 Mpc scales~\cite{magCMBR} (an exception is Ref.~\cite{JKO00}). 
Such fields can, however, realistically only
be produced during an inflationary scenario, with the stringent requirements
of breaking conformal invariance and avoiding backreaction
of the created magnetic fields on the inflationary process. To be observable,
field strength of $B\sim 10^{-9}$ Gauss~\cite{remark} have to be assumed. 
Field of that strength may, however, already potentially be ruled out due 
to likely overproduction of magnetic fields in galaxy clusters~\cite{BJ04}. 

When magnetogenesis happens after inflation, resulting magnetic field
spectra are blue, with much more power on small scales than on large scales.
For dynamically relaxed magnetic fields a correlation between the final
present day magnetic field strength $B$ and its correlation length $L$ may
be given~\cite{BJ04}
\begin{equation}
B\simeq 5\times 10^{-12}{\rm Gauss}\biggl(\frac{L}{{\rm kpc}}\biggr)
\end{equation}
On larger scales fields are likely falling of with a white noise spectrum
$B\sim (l/L)^{-3/2}$~\cite{JS10}
or even steeper~\cite{CD03}. 
Magnetic fields on kpc scales are
usually not believed to change the observable anisotropies in the CMBR
since that scale would correspond to multipoles of
$l\sim 10^7$ whereas the Planck mission will observe only 
up to $l\sim 2-3\times 10^3$. We will show here that this view is incorrect, 
i.e. magnetic fields on such small scales {\it do} 
change the anisotropies in the CMBR on smaller multipoles. 

Shortly before recombination
CMBR photons do not participate in fluid flows on kpc scales, as the
photon mean free path is much larger $l_{\gamma}\sim $ Mpc.
They do, however, strongly affect fluid flows by introducing a high 
drag on moving electrons due to occasional Thomson
scatterings, leaving the plasma on small scales
in a highly viscous state~\cite{JKO98} before recombination. Immediately
after the decoupling of photons on scale $L$
(i.e. when $l_{\gamma}$ becomes larger than $L$) the plasma experiences
an enormous $\sim 3\times 10^{-5}$
decrease in the speed of sound from $c_S = 1/\sqrt{3(1+R)}$
where $R = 3\rho_b/4\rho_{\gamma}$ with $\rho_b$, $\rho_{\gamma}$ the
photon and baryon mass densities, respectively, to 
$c_s = \sqrt{2T/m_b}$~\cite{remark2}. That is, whereas for all purposes
the plasma had been incompressible when $l_{\gamma}<L$ it becomes 
compressible, at least for sufficiently large magnetic field strength, when
$l_{\gamma}>L$. 

Imagine a stochastic magnetic field and negligible velocities $\bf v$ initially. The evolution of velocities and densities are given by the Euler and continuity
equations 
\begin{eqnarray}
\label{eq:fluid1}
\frac{d\bf v}{d t} + \bigl({\bf v}\cdot {\bf\nabla}\bigr)\cdot{\bf v}
+ c_s^2\frac{{\bf\nabla} \rho}{\rho} & = & - \alpha {\bf v}
-\frac{1}{4\pi\rho}{\bf B}\times
\bigl({\bf \nabla}\times {\bf B}\bigr)\\
\label{eq:fluid2}
\frac{d\rho}{d t} + {\bf \nabla}\bigl(\rho {\bf v}\bigr) & = & 0  
\end{eqnarray}
where $\alpha\sim 1/l_{\gamma}$ (cf.~\cite{BJ04}) is the photon drag term. 
In the overdamped, highly viscous state before recombination, 
only the terms on the
RHS of Eq.~(\ref{eq:fluid1}) are important. Very quickly 
($\Delta t\sim 1/\alpha$) terminal velocities of 
$v\simeq v_A^2/\alpha L$ are reached. 
Here $v_A = B/\sqrt{4\pi\rho}$ is the Alfven
velocity of the baryon plasma. For a stochastic field the
generated fluid flows are necessarily both rotational 
(i.e. $\nabla\times {\bf v}\neq 0$) and compressional 
(i.e. $\nabla\cdot{\bf v}\neq 0$). In fact, one can show that the energy
dissipation rate of compressional modes is a factor $(v_A/c_s)^2$ smaller
than those for rotational modes, such that when $c_s\gg v_A$ a larger part
of magnetic field configuration which induce compressional flows 
could potentially survive.
The compressional component leads to the creation of density
fluctuations. Using Eq.~(\ref{eq:fluid2}) one finds 
$\delta\rho/\rho (t)\simeq v t/L\simeq v_A^2 t/\alpha L^2$.  
These density fluctuations become larger with time until either, pressure
forces become important in counteracting further compression,
or the source magnetic stress term decays. The former happens when the
last term on the LHS of Eq.~(\ref{eq:fluid1}) 
$(c_S^2/L)\,\delta\rho/\rho$  is of the
order of the magnetic force term $v_A^2/L$. That is, density
fluctuations may not become larger than 
$\delta\rho/\rho \lsim (v_A/c_s)^2$. Here Alfven- and sound speed shortly
before recombination are given by 
\begin{eqnarray}
v_A & = & \, 4.59\, \frac{\rm km}{\rm s}\,\biggl(\frac{B}{3\times 10^{-11}{\rm Gauss}}\biggr)\biggl(\frac{T}{0.259 {\rm eV}}\biggr)^{1/2} \\
c_s & = & \, 8.47\, \frac{\rm km}{\rm s}\,\biggl(\frac{T}{0.259 {\rm eV}}\biggr)^{1/2} 
\end{eqnarray} 
It has been shown in Ref.~\cite{BJ04} that magnetic fields do decay even in the
viscous photon free-streaming limit applicable shortly before recombination. 
Here decay of magnetic energy occurs via the excitation of fluid flows 
which are than converted to heat due to photon drag. Though counterintuitive,
dissipation is stronger when the drag term $\alpha$ becomes weaker.
By direct numerical simulation the linear analysis~\cite{JKO98} and non-linear
estimate~\cite{Subra98} 
was confirmed that magnetic fields do decay when the eddy 
turnover rate
$v/L\simeq v_A^2/\alpha L^2$ equals the Hubble rate $H\simeq 1/t$.
Entering this into the above expression for $\delta\rho/\rho (t)$ one finds
that the average density fluctuation is not expected to exceed unity 
by much, even for vanishing $c_s$. Putting all this together, we expect
\begin{equation}
\frac{\delta\rho}{\rho}\simeq {\rm min}\,
\biggl[1,\biggl(\frac{v_A}{c_s}\biggr)^2\biggr]
\end{equation} 
for the density fluctuations generated by magnetic fields before
recombination.

The cosmological hydrogen recombination process is well approximated 
by the following differential equation in time~\cite{Peebles}
\begin{eqnarray}
\label{eq:recomb}
\frac{{\rm dn_e }}{{\rm d} t} + 3Hn_e & = & -C\biggl(
\alpha_en_e^2-\beta_en_{H^0}
{\rm e}^{-h\nu_{\alpha}/T}\biggr) \\
C & = & \frac{1+K\Lambda n_{H^0}}{1+K(\Lambda +\beta_e) 
n_{H^0}}
\end{eqnarray}
where $n_e$, $n_{H^0}$, and $n_H$ are electron-, neutral hydrogen- and
total hydrogen- $n_H=n_{H^0}+n_e$ density
respectively, and with $\alpha_e$, $\beta_e$, and $\Lambda$ the 
Case B recombination rate, photoionization rate from the $2s$ level, and the
$2s\to 1s$ two photon decay rate, respectively. Furthermore 
$h\nu_{\alpha}$ is the Lyman-$\alpha$ transition energy, $T$ is temperature,
and $K = \lambda_{\alpha}^3/(8\pi H)$ with $\lambda_{\alpha}$ the
Lyman-$\alpha$ wavelength and $H$ the Hubble constant. Note that 
Eq.~(\ref{eq:recomb}) is only for illustrative purposes as it
neglects the presence of helium.

\begin{figure}
\epsfxsize=8.5cm
\epsffile[0 0 600 530]{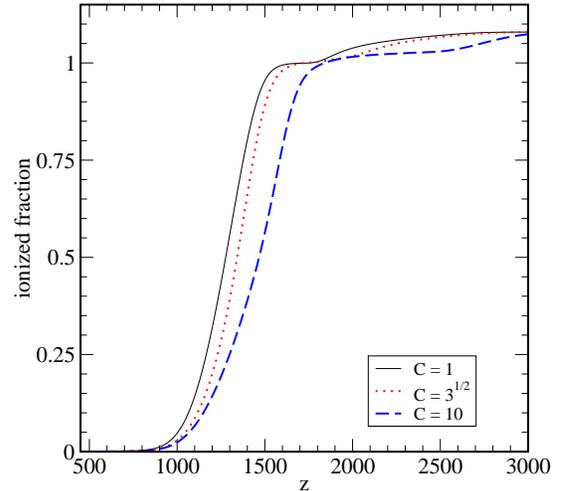}
\caption{Ionization fraction $\bar{X_e}$ for a Universe with the
preferred WMAP7-year cosmological parameters~cite{WMAP} 
as a function of redshift
for a homogeneous Universe (solid) and two Universes with small-scale
inhomogeneities $\sqrt{\langle(\delta n/n)^2\rangle} = \sqrt{3}$ 
(dotted) and $10$
(dashed), respectively.
Here, somewhat arbitrarily, the parameters $f_V^i = (0.567, 0.33, 0.1)$
and $\Delta^i = (0.1,1,6)$ for the first inhomogeneous model and
$f_V^i = (0.663, 0.33, 0.04)$ and $\Delta^i = (0.1,1,168)$
for the second model have been chosen, where $f_V^i$ is the
volume fraction filled with density $n^i = (1+\Delta^i)\langle n\rangle$
of zone $i$.}
\label{fig1}
\end{figure}

A key observation in Eq.~(\ref{eq:recomb}) is that it is non-linear in 
density in the first term on the RHS and in the factor $C$~\cite{remark3}.
In an Universe inhomogeneous on scales $L\ll l_{\gamma}$ CMBR anisotropies
depend on the average electron density $\langle n_e\rangle$.
However, due to the non-linearity 
$\langle n_e\rangle \neq n_e^{\rm homo}$ where $n_e^{\rm homo}$ is
the electron density in a homogeneous
Universe, irrespective of the fact that the
baryon density $n_b^{\rm homo}=\langle n_b\rangle$ equals the
average baryon density in the inhomogeneous Universe. This may be seen
in Fig. 1 where we computed with help of the public code 
RECFAST~\cite{Seager:1999km}
the ionization fraction 
$\bar{X_e}=\langle n_e\rangle/\langle n_H\rangle \neq \langle X_e\rangle$
in a inhomogeneous Universe by taking the average of the electron
densities of three independent regions~\cite{remark4} 
with different baryonic densities but with the same average density
as a homogeneous Universe. It is seen that the drop in $\bar{X_e}$, i.e.
recombination, occurs earlier when inhomogeneities exist.

CMBR temperature anisotropies may be calculated at linear
order by evolving
temperature $\Theta_0$ and gravitational potential $\Psi$
perturbations
of wavevector $k$ across the epoch of recombination.
When this is done the observed temperature fluctuations are related to
$\Theta_0 + \Psi = ({\hat\Theta}_0 +\Psi){\cal D}(k)$~\cite{CMBR}
where
\begin{equation}
{\cal D}(k) = \int_0^{\eta_0}{\rm d}\eta 
\frac{{\rm d} \tau}{{\rm d} \eta}\,{\rm e}^{-\tau(\eta ,\eta_0)}
{\rm e}^{\bigl[k/k_D(\eta )\bigr]^2}
\end{equation}
is due to the imperfect coupling of photons and baryons inducing
exponential damping of perturbations on the 
photon diffusion scale $k_D(\eta )$, and 
$g({\eta}){\rm d}\eta = {\rm d}\tau/{\rm d}\eta\,{\rm exp}(-\tau ){\rm d}\eta$
is the fraction of photons observed today (i.e. at present conformal
time $\eta_0$)
which scattered last between conformal times
$\eta$ and $\eta +{\rm d}\eta$, with $g({\eta})$ the visibility function.
Here ${\rm d}\tau/{\rm d}\eta = X_en_H\sigma_{Th}a$ with $\sigma_{Th}$ the
Thomson cross section and $a$ scale factor, such that $\tau$ is the photon optical depth. The damping factor ${\cal D}(k)$ strongly modifies the 
undamped ${\hat\Theta}_0+\Psi$ temperature fluctuations. The behavior of
${\hat\Theta}_0$ is given by the solutions of the equation of a forced 
oscillator. It is well known that due to well specified initial conditions (i.e. only growing modes) ${\hat\Theta}_0(k)$ exhibits an oscillatory
behavior with peaks given by $k_mr_s \simeq m \pi$ where $m$ 
is an integer. These peaks are due to perturbations having performed
half, one, one-and-a-half, ... sonic oscillations where
\begin{equation}
\label{shorizon}
r_S = \int_0^{\eta_{\rm rec}}c_s\,{\rm d}\eta = \int_{z_{\rm rec}}^{\infty}
c_s\frac{{\rm d}z}{H(z)}
\end{equation} 
is the sound horizon, with $\eta_{\rm rec}$ ($z_{\rm d rec}$)
conformal time (redshift) at recombination,
$c_s$ the baryon-photon speed of sound, and $H$ the Hubble constant. Corresponding peaks in the
temperature-temperature correlation function on angular
scale, or equivalently on spherical harmonic multipole $l$, are observed
at $l_m\simeq k_m(\eta_0-\eta_{\rm rec})$. The above gives us most of the ingredients to qualitatively understand modifications in the CMBR anisotropies 
from small scale inhomogeneity.

\begin{figure}
\epsfxsize=9.5cm
\epsffile[0 0 600 530]{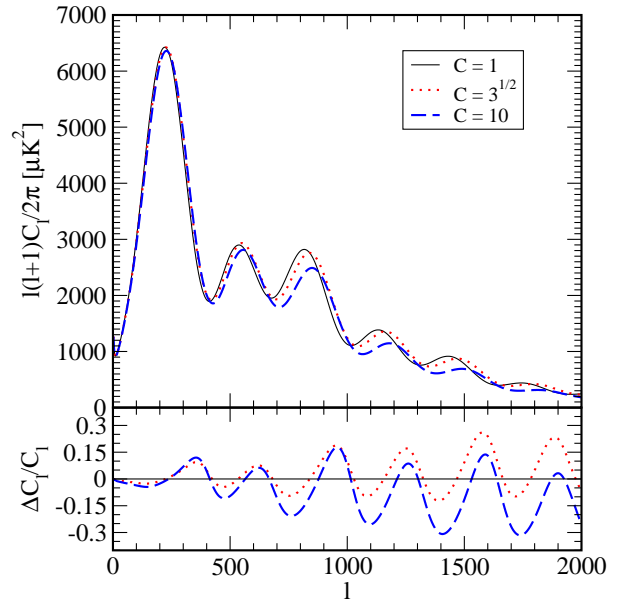}
\caption{CMBR anisotropies in conventional units (cf.~\cite{WMAP})
(upper panel) as a function of multipole for the best-fit WMAP7 year model,
and the two inhomogeneous models described in the caption of
Fig.~\ref{fig1}, as well as their fractional
differences (lower panel) to the best-fit WMAP7 year model.
The same line coding as in Fig.~\ref{fig1} is applied.}
\label{fig2}
\end{figure}

We modified CAMB~\cite{CAMB} to compute
the CMBR anisotropies when magnetic field induced baryon density
fluctuations are present. The results are shown in Fig.~\ref{fig2}
showing the anisotropies and their fractional deviations
from the best-fit WMAP7 model for the two inhomogeneous Universes with
$X_e$ shown in Fig.~\ref{fig1}. It is seen that
inhomogeneities have two main effects
(a) they move the Doppler peak locations to higher multipoles and (b) they enhance Silk damping of the high $l$ peaks. Both may be understood by 
inspecting Fig.~(\ref{fig1}). In small-scale inhomogeneous Universes
high density regions recombine earlier, making the average ionization
fraction $X_e$ drop significantly earlier, and therefore increasing the
redshift of ionization $z_{\rm rec}$. Low density regions which recombine 
later are not too important for the visibility function since they do not
contain too many electrons. For example, for 
$\sqrt{\langle(\delta n/n)^2\rangle} = \sqrt{3}$ recombination 
(the peak of the visibility function) is moved
from $z_{\rm rec}\approx 1078$ to $z_{\rm rec}\approx 1118$ a substantial
change of $6.6\%$. In order to induce such a large change of $z_{\rm rec}$
by a change of
the baryonic- or matter- densities, relative changes of $\sim 20\%$, 
$\sim 10\%$, respectively are required. These are
far beyond the WMAP7 and baryonic acoustic oscillation (BAO)~\cite{BAO} 
error bars on their respective values of $\sim \pm 2\%$, and $\sim\pm 3.3\%$,
respectively. An earlier recombination leads to Doppler peaks moving to
higher $l$, e.g. for $\sqrt{\langle(\delta n/n)^2\rangle} = \sqrt{3}$
all lower peaks are moved by $\sim 2.5\%$. This may be understood 
since to lowest order $\Delta l \simeq\Delta z_{rec}^{1/2}$ (see below).

The second effect, enhanced Silk damping, is somewhat more surprising since
the Silk damping scale is the diffusion scale (i.e. $d_{\gamma}\simeq
\sqrt{l_{\gamma}t}$ with $l_{\gamma}$ photon mean free path and $t$ time)
at recombination. Earlier recombination would imply less time for photon
diffusion and so less Silk damping. However, inspecting again 
Fig.~(\ref{fig1}) one observes that in the inhomogeneous Universes 
$\langle X_e\rangle$ is
smaller by $\sim 10\%$ already some time before recombination. This is due
to earlier helium recombination in the high density regions. A smaller
electron density implies larger $l_{\gamma}$ and therefore larger 
$d_{\gamma}$. As this latter effect dominates, the combined effect is more
Silk damping as evident from Fig.~(\ref{fig2}).

Fractional changes in the CMBR anisotropies in inhomogeneous Universes compared
to homogeneous Universes are substantial $\sim 10\%$ even for
$\sqrt{\langle(\delta n/n)^2\rangle}\sim 1$, particular at high multipoles.
On first sight one would think that such large changes may be detected or
ruled out by the Planck mission. To good approximation $l_m\simeq
k_m\eta_0\sim \eta_0/r_S\sim \eta_0/\eta_{\rm rec}\sim
[H_{\rm rec}(1+z_0)]/[H_0(1+z_{\rm rec})]$.
Assume for the moment that
the Universe only contains matter $\Omega_Mh^2=\Omega_{d}h^2+\Omega_bh^2$
and radiation, 
where $\Omega_{d}$, $\Omega_b$ are dark matter- and baryon- contributions
to the critical density today, and $h$ is the Hubble constant in units
of $100 {\rm km\, s^{-1} Mpc^{-1}}$. If one then assumes that the Universe
is critically closed $l_m$ becomes a function of only $z_{\rm rec}$
(since $[H_{\rm rec}(1+z_0)]/[H_0(1+z_{\rm rec})]
\sim z_{\rm rec}^{1/2}$ up to 
calculable radiation contributions)
independent of $\Omega_M$ or the Hubble constant. Since $z_{\rm rec}$
is only logarithmically dependent on well-know atomic physics, magnetic
field induced density fluctuations could be detected/ruled out to very
high precision via the shift of the Doppler peaks. 
Unfortunately the situation is somewhat more complicated
as the present Universe is dominated by a cosmological constant. In that
case $1/H_0\sim \eta_0 = f_1(\Omega_Mh^2, \Omega_{\Lambda}h^2)$ and
$H_{\rm rec} = f_2(\Omega_Mh^2, z_{\rm rec})$ (using that the radiation
density is well known), where $f_i$ are functions. To break possible degeneracies between density inhomogeneities and other cosmological
parameters $\Omega_Mh^2$ and $\Omega_{\Lambda}h^2$ must be known
accurately. Here $\Omega_Mh^2$ can be inferred from the CMBR anisotropies
and $\Omega_{\Lambda}h^2)$ from either supernovae surveys or BAO observations
of the angular diameter distance. Nevertheless, it is not clear if one can
achieve the desired accuracy. Alternatively, assuming a closed Universe
and using a precise measurement of the Hubble constant $h$ could also lead
to a fairly precise prediction of $z_{\rm rec}$ and the $l_m$'s. 
A more detailed analysis is beyond the scope of this letter and is deferred 
to a future publication.

Which field strengths are detectable in case the Planck mission combined with other surveys will be able to establish (or refute) the existence of small-scale inhomogeneity before recombination~\cite{remark6} ?
Following Ref.~\cite{BJ04} primordial magnetic fields do decay on the scale
implicitly given by $v/L\simeq H$, with $v\simeq v_A^2/L\alpha$ 
before recombination
in the viscous regime and $v\simeq v_A$ after in the turbulent regime.
Assuming that the initial spectrum is given by 
$B\sim L^{-n/2}$ one may deduce that magnetic field strength 
shortly before, $B^<$ and after $B^>$ recombination (which is also the
final present day field strength~\cite{BJ04}) are related by
$B^<\simeq (\alpha/H_{\rm rec})^{n/(2n+4)}B^>$. 
That is, due to the rapid disappearance
of photon drag during recombination, substantial amounts of magnetic field
energy density dissipates right at recombination since 
$\alpha/H_{\rm rec}\approx 170$. 
It is not clear, but subject to further investigation, if shocks resulting
from the magnetic stress acceleration and the accompanying 
shock ionization during recombination are also of importance. In any case, for a final magnetic
field strength of $B^>\simeq 10^{-11}{\rm Gauss}$ and a white
noise spectral index $n=3$ one finds $B^<\simeq 4.7\times 10^{-11}$Gauss which
comes close to fulfilling $v_A\sim c_s$ such that 
it should be potentially detectable.

In summary, we have argued that primordial magnetic fields induce 
small-scale baryon inhomogeneity of substantial amplitude. Due to non-linearities in the recombination equations such inhomogeneity induces 
changes in the ionization fraction before recombination. This, in turn,
influences the anisotropies in the CMBR, such that present day
primordial magnetic fields of strength $B\sim 10^{-11}$ Gauss could 
potentially be detected by a combination of future and present
CMBR, BAO, and supernovae observations.

\vskip 0.15in
{\it Acknowledgments}
We acknowledge a useful discussion with Levon Pogosian.


\begin{thebibliography}{99}
\bibitem{remark}{All cited length scales and magnetic field strengths
in this paper are comoving, i.e. the values
they would have at the present epoch are given.}
\bibitem{remark2}{Due to efficient CMBR cooling the effective speed of
sound is that for an isothermal baryon gas.}
\bibitem{remark3}{In fact, the nonlinearity in $C$ is only very mild
since $C$ may be well approximated by $C\approx \Lambda/(\Lambda + \beta_e)$
signifying that double photon decay is about one 
order of magnitude more important than Lyman-$\alpha$ transitions in
producing neutral hydrogen in the ground state.}
\bibitem{remark4}{Though the Lyman-$\alpha$ photon mean free path
is coincidentally of the order of $\sim 1$ kpc, i.e. 
different dense regions are not completely
decoupled from each other, the ``independent region'' 
approximation is very good since hydrogen formation by 
Lyman-$\alpha$ transitions is subdominate (cf.~\cite{remark3}).}
\bibitem{remark6}{There are no other plausible candidats for re-creating
small-scale inhomogeneity after Silk damping
which is volume filling and of magnitude $\delta\rho/\rho\sim 1$
known to the authors.}
\end{thebibliography}
\end{document}